\documentclass [11 pt, a4paper] {article}
\pdfoutput=1
\usepackage {jcappub}
\usepackage [font = small]{caption}
\usepackage {subcaption}

\title{Optimizing BAO measurements with 
non-linear transformations of the Lyman-$\alpha$ forest}

\author[1, 2]{ Xinkang Wang,}
\author[2] {Andreu Font-Ribera}
\author[1, 2] {and Uro\v{s} Seljak}

\affiliation[1]{Department of Physics, University of California, 
Berkeley, United States}
\affiliation[2] {Lawrence Berkeley National Laboratory, 
Berkeley, United States}

\emailAdd{xinkang.wang@berkeley.edu}
\emailAdd{afont@lbl.gov}
\emailAdd{useljak@berkeley.edu}

\newcommand{\lyaf}{Ly$\alpha$ forest}
\newcommand{\lya}{Ly$\alpha$}

\newcommand{\hMpcc}{\ h^{-3}\text{Mpc}^3}
\newcommand{\ihMpc}{\ h\text{Mpc}^{-1}}
\newcommand{\vk}{\textbf{k}}
\newcommand{\vC}{\textbf{C}}
\newcommand{\vPo}{\mathbf{P_o}}
\newcommand{\vPt}{\mathbf{P_t}}
\newcommand{\vkperp}{\mathbf{k_\perp}}

\abstract{We explore the effect of applying a non-linear transformation to the 
Lyman-$\alpha$ forest transmitted flux $F=e^{-\tau}$ and the ability of analytic 
models to predict the resulting clustering amplitude. 
Both the large-scale bias of the transformed field (signal) and the amplitude 
of small scale fluctuations (noise) can be arbitrarily modified, but we were 
unable to find a transformation that increases significantly the signal-to-noise
ratio on large scales using Taylor expansion up to the third order. In particular, however,
we achieve a 33\% improvement in signal to noise for \textit{Gaussianized} field in transverse direction.  
On the other hand, we explore an analytic model for the large-scale 
biasing of the \lyaf, and present an extension of this model to 
describe the biasing of the transformed fields. 
Using hydrodynamic simulations we show that the model works best
to describe the biasing with respect to velocity gradients, but is less
successful in predicting the biasing with respect to large-scale density fluctuations, especially 
for very nonlinear transformations.}


\begin {document}
\maketitle
\flushbottom

\captionsetup{labelsep = period}

\section {Introduction}
\label{sec:1}

One of the main drivers of current cosmological observations is the study of
the accelerated expansion of the Universe \cite {Weinberg 2012}. 
One particular probe of the acceleration
has recently proven very successful: 
the measurement
of the Baryon Acoustic Oscillation (BAO) scale in the clustering of galaxies,
which can be used as a cosmic ruler to study the geometry of the Universe
at different redshifts \cite {Seo 2003}. Even though the first measurements of 
the BAO scale were presented in the 
context of galaxy clustering at low redshift \cite {Eisenstein  2005, 
Cole 2005, Blake 2011, Anderson 2014}, in theory any tracer of the 
large-scale distribution of matter could be used to measure BAO. 
At high redshift ($z>2$), for instance, the BAO scale was recently measured in the 
clustering of the absorption features in distant quasar spectra 
\cite {Busca 2013, Slosar 2013, Ribera 2014, Delubac 2014}, a technique known as the Lyman-$\alpha$ forest, or \lyaf. 

As light emitted from distant quasars travels through an expanding Universe,
it is gradually redshifted towards longer wavelengths.
Photons that have escaped from the quasars with energy above the \lya\ transition energy might
be absorbed by neutral hydrogen at positions where the
redshifted wavelength of the photons equals that of the \lya\ transition energy.
Hence, photons emitted with different energies from a quasar will have to travel through different distances 
to be absorbed (when they reach the Ly$\alpha$ transition wavelength), and as a consequence the fraction of
transmitted flux for a given photon wavelength is a function of the neutral hydrogen density,
 which in turn is a function of line-of-sight position from the quasar
\cite {Rauch 1998, Meiksin 2009}. 
Although the distribution of neutral hydrogen on very small scales (e.g., a few kpc) 
is determined by the thermal properties of intergalactic medium, on
cosmological scales it can be described as a tracer of the underlying
density field, which was confirmed in the first studies of the clustering of the absorption
features in high resolution spectra from a handful of bright quasars
\cite {Croft 1998, Croft 1999, McDonald 2000}.  
Since then, \lyaf\ has become an increasingly important tool
to study the large-scale structure of the Universe,
and up to now it is considered one of the most promising methods to study the
clustering on both the smallest scales (via the correlated absorption along a
single line of sight) and on the largest scales (via the correlations in
neighboring lines of sight). 

An interesting question that we will address in this paper is whether the information extracted from a 
\lyaf\ survey can 
be increased by a nonlinear transformation of the observed field. 
Similar approaches have been suggested in the context of weak lensing and 
galaxy clustering \cite {Neyrinck 2009, Seo 2011, Joachimi 2011, Carron 2014}.
In \cite{Seljak 2012}, 
it was suggested that one could increase the signal-to-noise ratio of a measurement
by applying proper non-linear transformation to the \lya\ transmitted flux 
fraction (i.e.,  $F \rightarrow g(F)$). 
In this paper, in particular, we study whether such transformations could improve the
BAO measurement from the \lyaf.
We will also compare the analytic predictions of \lyaf\ bias in 
\cite{Seljak 2012} (for both the flux field and the transformed fields) to simulations. 

Specifically, the performance of a \lyaf\ BAO survey can be quantified by the error bars in 
the measurements of the BAO scale along and across the line of sight, respectively,
as well as their correlation coefficient. 
In order to optimize this measurement, we would like to have realistic 
simulations of a hypothetical survey and directly study the effect of different
transformations on the error bars of the measured BAO scales. 
Unfortunately, this is beyond the reach of current computational facilities. 
Hydrodynamic simulations are thus required in order to accurately reproduce 
the statistics of the \lyaf. However, these simulations are not able to 
simulate boxes large
enough to cover a full BAO survey. On the other hand, 
in order to test the analysis pipeline and study the effect of potential 
systematics, current BAO analyses using the \lyaf\ rely on simplified mocks \cite {Ribera 2012, Bautista 2014}. 
Despite the fact that these mocks have
the correct 1- and 2-point correlation functions, their higher order statistics are generally not 
correct, and thus would not be useful for our study.   

As a forecast, in section \ref{sec:sn} we will show that under several assumptions the effect of 
a non-linear transformation  $F \rightarrow  g(F)$ 
boils down to a single quantity: 
the ratio of large-scale biasing of the transformed field (squared) 
over the one-dimensional power spectrum of the field in large-scale limit, both of which
can be measured with current hydrodynamic simulations, 
allowing us to study the effect of non-linear transformations on the \lyaf\
in the context BAO measurements. We present the results of this study 
in section \ref{sec:trans}.  In addition, in order to better understand the effect of the transformations, in 
section \ref{sec:model} we extend the bias model of \cite{Seljak 2012} to 
describe the bias of non-linear analytic transformations. 
We conclude our results in section \ref{sec:conc}.

\section{Signal-noise ratio (S/N) in a BAO measurement}
\label{sec:sn}

The main goal of a BAO survey is to provide a measurement of the BAO 
scale at a given redshfit, as precise as possible. 
In order to predict the performance of a BAO survey, as a result, we would like
to generate simulations of the survey, measure its power spectrum, fit the BAO
scale on this measurement and look at its precision. 
We would like the simulation to be as realistic as possible, and we would 
like the analysis to be similar to that used with the real data.

In order to simulate accurately the \lya, however, we need to use 
expensive hydrodynamic simulations with a really high spatial resolution \cite {Lukic 2014}. 
This sets the maximum volume that one can simulate to roughly $100^3 \hMpcc$,
clearly insufficient to measure the BAO scale. Fortunately, there are several
approximations that we can use to make the problem more treatable.
Specifically, if the detection of the BAO scale is statistically very significant, 
the variance of the BAO measurement will scale proportionally to the 
covariance matrix of our power spectrum measurement, relative to the signal
level (the power spectrum itself).  
Moreover, assuming that linear theory holds on the relevant scales, the 
different bins in the power spectrum can be treated as independent, resulting
in a diagonal covariance matrix.

We can therefore quantify the precision of a given survey with 
a signal-to-noise function $S/N(\vk)$, defined as the ratio of the power
spectrum (signal) and its uncertainty (noise). 
In this section, we will present further approximations that will allow us
to compress the information of this function into a single number.

\subsection{Signal: large scale biasing}

The BAO information in a \lyaf\ survey is contained in Fourier modes with 
rather small wave numbers $k<0.3 \ihMpc$, where we expect the power spectrum
$P_F(\vk)$ to follow a simple linear bias model:
\begin{equation}
 P_F(\vk) = b_F^2(\mu) ~ P_m(k) ~,
 \label{eq:pf}
\end{equation}
where $P_m(k)$ is the matter power spectrum, $\mu = \hat {k} \cdot \hat{z}$ 
with $ \hat {z}$ being the line-of-sight direction, and $b_F(\mu)$ describes
the large-scale biasing of the \lyaf\ including redshift space distortions.
The angular dependence of the biasing follows the usual Kaiser formula
\cite{Kaiser 1987}:
\begin{equation}
 b_F(\mu) = b^\delta_F + (f\mu^{2} ) b^\eta_F, 
 \label{eq:bf}
\end{equation}
where $f$ is the logarithmic growth rate. In particular, this growth rate is close to one  in a 
$\Lambda$CDM universe at the redshifts of interest ($z>2$), where the universe
is close to Einstein-de Sitter, and will taken as unity from now on except stated otherwise. 
$b_F^\delta$ and $b_F^\eta$ are related to the response of the transmitted flux
fraction $F$ to small variations of the density and of the line-of-sight
velocity gradient, respectively. 
In section \ref{sec:model} we will present an analytic model to predict their values,
but for now we can think of these as free parameters.

An analytic non-linear transformation of flux $F$ is an analytic function $g(F)$. 
Assuming linear theory holds in the relevant scales,
the power spectrum of this transformed field, $P_g(\vk)$, can be described using the same
 equations \ref{eq:pf} and \ref{eq:bf} 
with ($b_{F}^{\delta}, b_{F}^{\eta}$) replaced by a different set
of bias parameters: ($b^\delta_g$, $b^\eta_g$). 
Therefore, the impact of the transformation on the BAO signal can be 
described by the ratio of the bias parameters squared: $b_{g}^{2}(\mu) / b_{F}^{2}(\mu)$. 

Note that most studies of the large-scale clustering of the Ly$\alpha$ forest focus on the 
statistics of the fluctuations around the mean transmitted flux fraction, i.e., 
$\delta_F = F/ \left< F \right> - 1$. 
The transformed field, however, might have an arbitrarily small mean, and for 
simplicity we choose to focus on the clustering of the $F$ and $g(F)$ fields
themselves.

\subsection{Noise: 1D power spectrum}

The error bars in the BAO measurements are set by the uncertainty with which
we are able to measure the clustering on the Fourier modes relevant for BAO.
Following \cite {McDonald 2007}, the uncertainty on a given Fourier
mode is proportional to
\begin{equation}
 \sigma^2_{P_F}(\vk) = 2 \left( P_F(\vk) + P_N^{eff} 
	+ A ~ P^{1D}_F(k_\parallel) / n^{2D}_q \right)^2 ~.
 \label{eq:spf}
\end{equation}
There are three contributions to the noise: i) the first term, $P_F(\vk)$, is 
set by the signal and is usually referred as \textit{cosmic variance};
ii) the second term, $P^{eff}_N$, is set by the level of instrumental noise
in the spectra, and is usually referred as \textit{noise power};
iii) the last term, $A ~ P^{1D}_F(k_\parallel) / n^{2D}_q$, takes account of the fact
that we can only sample the universe in a set of thin lines of sight, and is thus
set both by the level of intrinsic fluctuations (or 1D power $P^{1D}(k_\parallel)$) 
on a scale $\sim$ $k_\parallel=k \mu$ and by the 2D density of background
sources $n^{2D}_q$ \footnote{The constant $A$ depends on the distribution of 
weights in the different spectra, but for a given survey this can be treated 
as a constant}. This last term is usually referred as \textit{aliasing noise}
and is equivalent to the \textit{shot noise} in galaxy clustering studies. 
Now, as noted in \cite {MW 2011},  
$P^{1D}_F(k_\parallel)$ is approximately flat on scales relevant for BAO analyses.
Since the noise power is also flat 
(white noise), the last two terms in eq.\ref{eq:spf} can thus be described by a 
single term $P^{1D}_F(k_0) / n_{eff}$, where $P^{1D}_F(k_0)$ describes the 
typical level of $P^{1D}_F(k_\parallel)$ in the flat regime (i.e., $k_{0}$ large enough) 
and $n_{eff}$ is an effective density of lines of sight that takes into account the 
distribution of noise power in the spectra.

In present and near-future 
\lyaf\ BAO surveys, the contribution from cosmic 
variance is rather small, and we will ignore it in this study. The other
two contributions are comparable, but for simplicity in this study we will 
only consider the aliasing term.
And since the instrumental noise is very close to Gaussian,
any non-linear  transformation applied to the observed flux field will make the noise 
properties non-Gaussian. 
By ignoring the noise in our analysis, we are simplifying the problem, and 
our results should be read as the most optimistic case in the limit of low noise. 
We will thus characterize the 
effect of the non-linear transformation 
on the BAO uncertainties with the ratio of the amplitudes of the 
one-dimensional power in small $k_\parallel$ limit:
$P^{1D}_g(k_0) / P^{1D}_F(k_0)$. 

After all, for any non-linear transformation $ F \rightarrow g(F)$, 
we expect the following proportionality holds on large scales:
\begin{equation}
S/N (\textbf{k}) \propto R_{g} (\mu) = b_{g}^{2}(\mu) / P_{g}^{1D} (k_{0}).
\label{prop}
\end{equation}
The right-hand side of this equation may serve as a proxy for S/N and 
is a measurable quantity in simulations. We will focus on the value of $R_{g}$  in the rest of the paper.

\subsection{Numerical simulations}
\label{ss:sims}

In next section we will study the effect of analytic non-linear transformations 
$g(F)$ on the signal-to-noise ratio (S/N) with which we can measure BAO. 
To do that we will use hydrodynamic simulations to measure the large-scale biasing (squared)
of the transformed field, $b_{g}^{2}(\mu)$ (proxy for signal), and to measure the 1D power 
spectrum of the transformed in low-$k$ limit, $P_g^{1D}(k_0)$ (proxy for noise).

The simulation we use is one of the Gadget-3 simulations adopted in the 
Nyx/Gadget-3 comparison project~\cite{Stark}, with a box size of 
$40 h^{-1}$Mpc and a total of $2 \times 1024^{3}$ particles. 
The simulation used a WMAP7 cosmology ($\Omega_b = 0.046$, 
$\Omega_m = 0.0275$, $\Omega_\Lambda = 0.725$, $h = 0.702$, $\sigma_8 = 0.816$,
and $n_s = 0.96$) and the ionizing background prescription of \cite{Gig 2009}; 
it also used the QUICKLYA option in Gadget to implement a simple star 
formation recipe.
We have rescaled the optical depth in the box in order to have a mean flux
of $\langle F \rangle = 0.8413 $ (at $z=2$), in agreement with estimates
from SDSS \cite{McDonald 2005}. In the main part of this paper we use only the simulation output at $z=2$, but
in appendix \ref{app:z} we also present results using outputs at redshift 
$z=2.4$ and  $z=2.75$. 

\begin{figure}
\includegraphics[width = 17cm, height = 13cm,keepaspectratio= true, trim = 2cm 17.8cm 0cm 3cm]{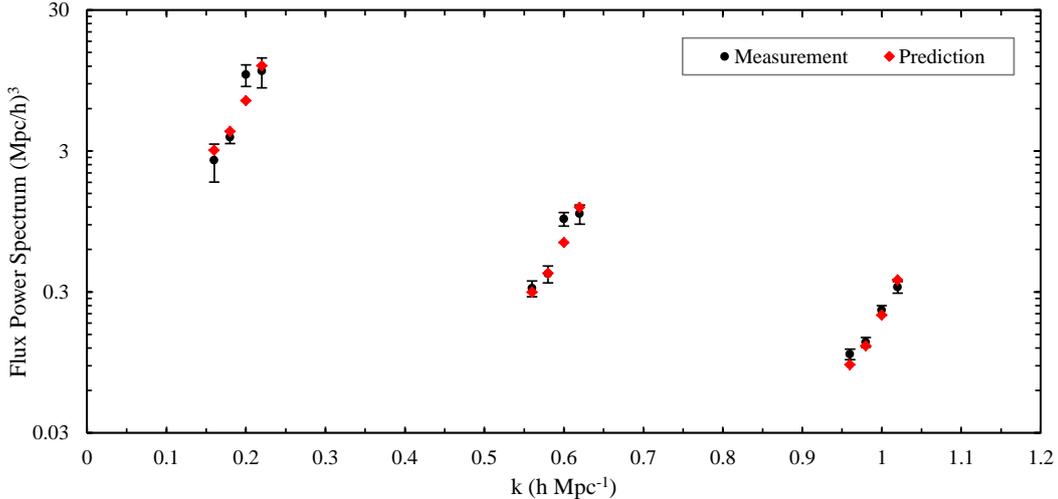}
\caption{The plotted scales are $k=0.2, 0.6, 1.0 h$ Mpc$^{-1}$, each of which 
has four $\mu$ bins. For clarity, at each $k$ the power spectrum with 
$\mu=0.125$ is plotted at $(k-0.04)$, $\mu=0.375$ is plotted at $(k-0.02)$, 
$\mu=0.625$ at $k$, and $\mu=0.875$ at $(k+0.02)$.}
\label{fig:p3d}

\end{figure}
\begin{figure}
\includegraphics[width = 17cm, height = 13cm, keepaspectratio= true, trim = 2cm 17.8cm 0cm 2.8cm]{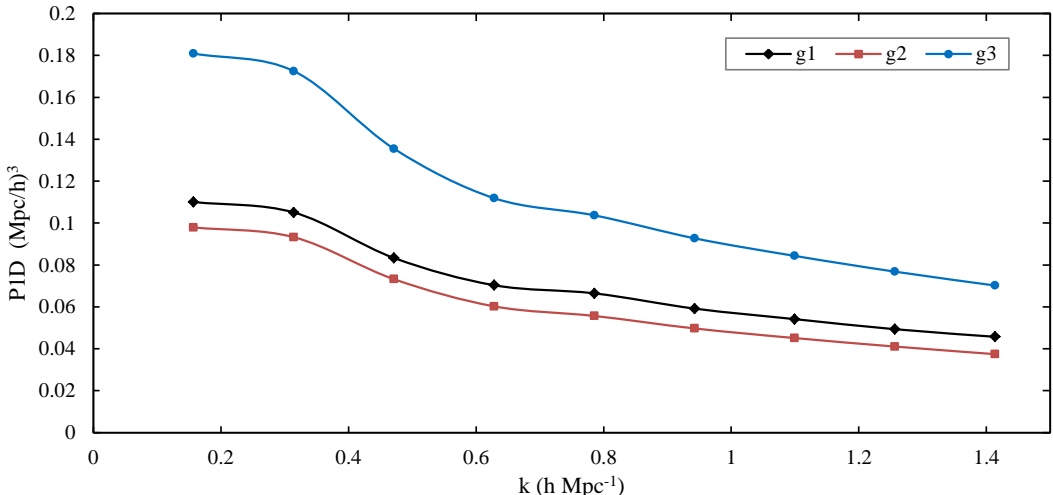}
\caption{1D power spectrum measured in simulations for the original field 
$g_1(F)=F$ (black line), and the two other fields discussed in the next 
section. We use the 1D power in the fundamental mode of the simulation 
($k_0=0.157 \ihMpc$) as a proxy for the noise in a BAO measurement.}
\label{fig:p1d}
\end{figure}

We now show that the proxies for signal and noise 
we chose in previous two subsections  (i.e., eq.\eqref{prop}) are valid, respectively. 
In figure \ref{fig:p3d}, we show (in black) the 3D power spectrum of the 
transmitted flux fraction $F$, measured at three different scales 
$k=0.2, 0.6, 1.0 \ihMpc$, and at four different line-of-sight directions 
($\mu=0.125, 0.375, 0.625, 0.875$). 
The red points show the prediction using eq.\ref{eq:pf} for the 
values of ($b_F^\delta=-0.0733$, $b^\eta_F=-0.0995$) that best fit the 
simulations (see appendix \ref{app:b} for a description of how we fit 
the bias parameters in simulations). The figure clearly shows the validity 
of squared bias as the proxy for signal. On the other hand, 
it is important to remind the reader that throughout this paper, we will \textit{not} 
follow the common convention of measuring statistics of fields that have been
normalized (i.e., divided by its mean). Therefore, if one wishes to compare the 
flux biases reported in this paper, one would have to divide them by the mean flux
in the simulation: $ \langle F \rangle =0.8413$.

In figure \ref{fig:p1d} we show the measured 1D power spectrum of the 
transmitted flux fraction $g_1(F)=F$, and two of the transformed fields 
discussed in next section: $g_2(F) = -0.261F+F^{2}$ and 
$g_3(F) = 1.084F - 3.039F^2 + F^3$. 
In the figure it is clear that $P_{g}^{1D}(k)$ is close to constant at scales
relevant in BAO analyses ($k<0.3\ihMpc$). 
We will use the 1D power at the fundamental mode, $k_0=0.1571\ihMpc$, as a 
proxy for the noise in a BAO measurement.

\section{S/N of non-linear transformations}
\label{sec:trans}

In the previous section we have shown that under several assumptions,
the performance of a BAO survey can be quantified using the ratio of the 
large-scale bias (squared) over the low $k$ limit of the one dimensional power 
spectrum. 
We also presented the measurement of these quantities in the original \lya\
absorption field using hydrodynamic simulations. 
In this section, we will use the same simulations to measure the equivalent
quantities in different fields which are obtained by performing non-linear 
transformations to the  original field.

\subsection{Gaussianized field}

\begin{figure}
 \centering
 \includegraphics [width = 17cm, height = 13cm, keepaspectratio= true, trim = 2cm 17.8cm 0cm 2.8cm]{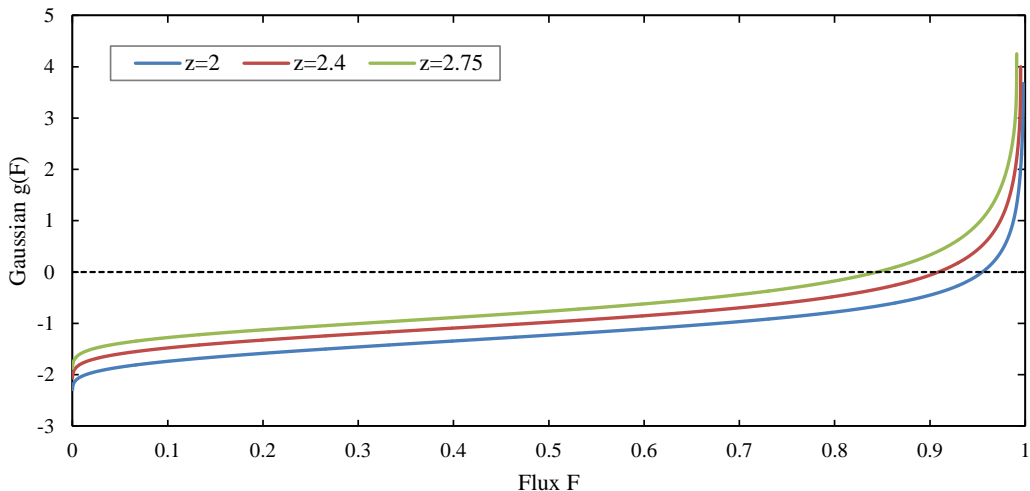}
 \caption{Monotonical relation $g(F)$ to convert the simulated flux field 
  into a Gaussianized field with zero mean and unit variance. 
  The different lines correspond to $z=2$ (blue), $z=2.4$ (red) and 
  $z=2.75$ (green). The dashed line (black) is the ``zero" reference line.}
 \label{fig:gF}
\end{figure}

A Gaussian field can be fully described by its 2-point correlation function. 
The \lyaf\ flux field, on the other hand, is highly non-Gaussian, implying that 
in order to obtain all of its cosmological information we would need to use 
higher order statistics. 
Therefore, it is tempting to transform the flux field to make it as Gaussian
as possible \cite {Weinberg 1992, Croft 1998}. For instance, we can assume a monotonic relation between our 
flux field $F$ and a \textit{Gaussianized} field $g(F)$, which is implicitly defined via
\begin {equation}
\int_{-\infty}^{g(F)} dg^\prime p_g(g^\prime) 
				= \int_{0}^{F} dF^\prime p_F(F^\prime) ~,
\label{eq:gF}
\end{equation}
where $p_g(g)$ and $p_F(F)$ are the probability distribution function (PDF) 
of each field.  After measuring the PDF of flux in simulations, $p_F(F)$
(shown in figure \ref{fig:pdf}), we are able to determine the monotonic 
function $g(F)$ and transform the flux into a Gaussianized field, with zero mean 
and unit variance. In figure \ref{fig:gF} we show the transformation 
measured at different redshifts. We apply such transformation to the flux 
field in simulation and measure the S/N (i.e., $R_{g} (\mu)$ in eq.\eqref{prop}) of the 
resulting Gaussianized field.

We measure the one-dimensional power of the Gaussianized field, $P^{1D}_g(k_0)$, to
be $19.20 \pm 0.05$ times higher than that of the original field $F$. 
On the other hand, the bias parameters of the Gaussianized field are also larger,
by a factor of $5.06 \pm 0.32$ for $b_{g}^{\delta}$ and a factor of $2.70 \pm 0.41$ 
for $b_{g}^{\eta}$. Interestingly, this result implies a very anisotropic gain in signal-to-noise
ratio: while the S/N transverse to the line of sight ($\mu=0$) is measured 
a factor of $1.33 \pm 0.17$ larger, the S/N along the line of sight ($\mu=1$) is 
measured a factor of $0.71 \pm 0.08$ smaller. In addition, in large scale structure analyses 
it is common to define a redshift space
distortion parameter $\beta = f b_\eta / b_\delta$. 
In our simulations we measure it to be roughly $\beta_F=1.30$ for the original
flux field $F$. The Gaussianized field $g(F)$ described above, however, has
weaker redshift space distortions with only $\beta_g = 0.69$, where we have 
used growth rate $f=0.96$.

Finally, since standard \lyaf\ BAO measurements preferentially measure the line-of-sight 
scale \cite {Busca 2013, Slosar 2013, Delubac 2014}, with
very large uncertainties in the transverse direction, the fact that the 
measurement in the transformed Gaussianized field favors the transverse direction 
might be of special interest.

\subsection{Generic analytic transformations}
\label{ss:Fn}

The ultimate goal of this paper is to find an analytic non-linear transformation that maximally enhances
S/N with respect to that of the orignal flux field. To achieve this goal, we need to consider the
 entire set of generic analytic transformations $g(F)$ which can be 
expressed as an infinite Maclaurin series:
\begin{equation}
 g(F) = \sum_{m=1}^\infty a_m ~ F^m ~, \quad a_{m} = \frac{g^{(m)} (0)}{m!}.
\end{equation}
 In practice, however, 
we will study only the transformations involving terms up to 
$F^n$  ($a_{n} \neq 0$), and refer to these as transformations of order $n$.
Notice that we have ingored the zeroth order  in the series, of which the specific value 
has no effect on the following study.

We can think of $F^m$ as another tracer of the density field, with bias 
parameters $b_{F^m}(\mu) = b^\delta_{F^m} + (f\mu^{2}) b^\eta_{F^m}$. 
In linear regime, the large-scale cross-correlation of two tracers is 
just proportional to the product of their bias parameters, and therefore 
the cross-power spectrum can be expressed as the geometric mean of their 
power spectra. 
This allows us to define the biasing of the transformed field $g(F)$:
\begin{equation}
 b_g(\mu)  = \sum_{m=1}^n a_m ~ b_{F^m}(\mu) = \sum_{m=1}^{n} a_{m} (b_{F^{m}}^{\delta} 
+ (f\mu^{2})b_{F^{m}}^{\eta}) =   b_{g}^{\delta} + (f\mu^{2})b_{g}^{\eta}.
\end{equation}

On the other hand, for the noise, 
the one-dimensional power spectrum $P^{1D}(k_\parallel)$ is related to the
three-dimensional power spectrum $P(k_\parallel,\vkperp)$ by an integral
over $\vkperp$. 
Even in the limit of small $k_\parallel$, the 1D power is affected by high 
$\vkperp$ modes that are not well described by linear theory.  
This implies that the 1D power is not directly proportional to the bias 
parameters, and that the 1D cross-power of two fields $P^{1D}_{(F^m,F^i)}$ 
can not described by the geometric mean of their 1D powers.  Therefore, 
if we are interested in studying generic transformations up 
to a certain order $n$, we have to measure not only all the 1D power spectra 
$P^{1D}_{F^j}(k_0)$ where $j \leq n$, 
but also all the 1D cross-power spectra 
$P^{1D}_{(F^j,F^k)}(k_0)$:
\begin {equation}
P_g^{1D}(k_0) = \sum_{m=2}^n \sum_{i=1}^{m-1} a_i ~ a_{m-i} 
	~ P^{1D}_{(F^i,F^{m-i})}(k_0),
\end{equation}
where $P_{(F^{i}, F^{m-i})}^{1D} = P_{F^{i}}^{1D}$ if $ i =m-i$. 
Notice that $k_{0}$ is the fundamental mode.

To compute S/N for $g(F)$ of order $n$, we can compute in simulations the relevant biases 
$b_{F^m}$ (for $m \leq n$) and 1D auto- and cross-power spectra 
$P^{1D}_{(F^m,F^i)}(k_0)$. Moreover, multiplying the field by a constant would 
have no effect on the signal-to-noise 
ratio, since both quantities would be modified by the same multiplicative factor. 
Therefore, we are only interested in $g(F)$ of the \textit{monic} form: 
$g(F) = q_{1}F + q_{2}F^{2} + \cdots + q_{n-1}F^{n-1} + F^{n}$, 
where $q_{i} = a_{i}/a_{n}$. 

Even though the signal is anisotropic, we can compute an angularly average 
signal (the power spectrum monopole), and use a single averaged bias parameter
to quantify the signal: 
\begin{equation}
 \bar{b}_{g}^{2} = \int_0^1 d\mu ~ b_{g}^{2}(\mu) = (b_{g}^{\delta})^2 
	+ \frac{2}{3} b_{g}^\delta b_{g}^\eta + \frac{1}{5} (b_{g}^{\eta})^2~.
\end{equation}

\noindent Finally, we have an \textit {averaged} version of S/N for any $n^{th}$ order generic transformation:
\begin{equation}
\bar{R}_{g} = \frac{\bar{b}_{g}^{2}}{P_g^{1D}(k_0)} = f(q_1, q_2, ..., q_{n-1}).
 \label{eq:sn}
\end{equation}
The functional form $f$ is to indicate there are only $n-1$ variables we need to consider for 
$n^{th}$ order $g(F)$. Most importantly, at any given order, by maximizing $\bar{R}_{g}$ we can find the 
transformation that should result in a larger gain in signal-to-noise ratio. 
We will compare the results to the value of $\bar{R}_{g}$ for the original 
flux field: $\bar{R}_{F}=0.1109$.

\subsubsection{Results at second order}

We consider an arbitrary quadratic field $g(F) = q_1 F + F^2$, which only contains a single
parameter $q_1$. Using the measured bias values 
($b^\delta_F$, $b^\eta_F$, $b^\delta_{F^2}$, $b^\eta_{F^2}$), 1D power 
($P^{1D}_F(k_0)$, $P^{1D}_{F^2}(k_0)$) and cross-power ($P^{1D}_{(F,F^2)}(k_0)$),
we can estimate the angularly averaged signal-to-noise ratio function, $f(q_1)$, and numerically/analytically
find the global maximum of this function.

\begin{figure}
 \centering
 \includegraphics[width = 17cm, height = 13cm, keepaspectratio= true, trim = 2cm 17.8cm 0cm 2.5cm]{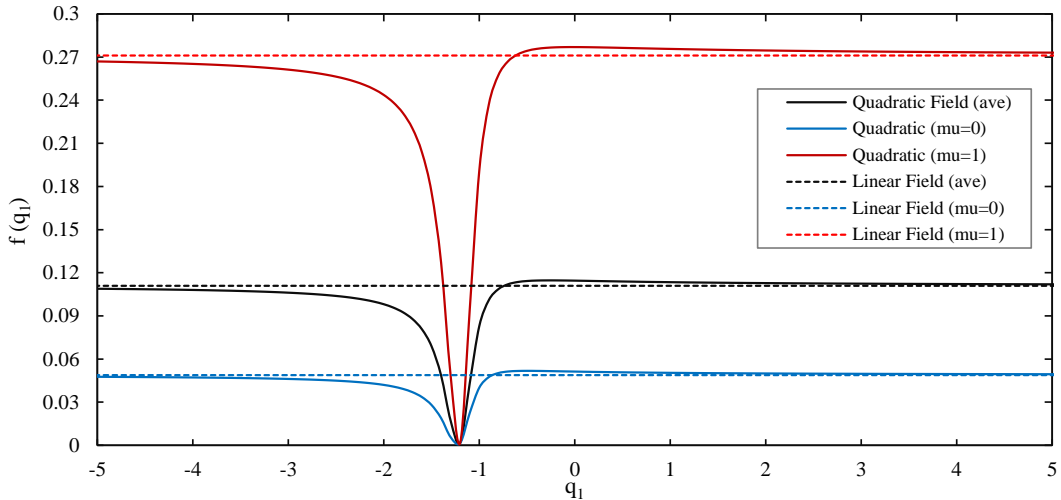}
 \caption{Signal-to-noise ratio for generic quadratic transformations
  $g(F)=q_1 F + F^2$, as a function of $q_1$. The line-of-sight direction
  ($\mu=1$) is shown in red, the transverse direction ($\mu=0$) is shown in 
  blue and the angular average (equation \ref{eq:sn}) is shown in black. 
  For large absolute values of $q_1$, we recover the signal-to-noise ratios 
  of the original field $F$ (dashed lines).}
 \label{fig:f2}
\end{figure}

The signal-to-noise ratio function $f(q_1)$ for a generic quadratic transformation is shown in 
figure \ref{fig:f2} (in black), in which two different angular configurations are also shown for comparison: 
$R_{g}(\mu=1)$ along the line of 
sight (in red), and $R_{g}(\mu=0)$ across the transverse direction (in blue).
As expected, for large absolute values of $q_1$ we recover the 
signal-to-noise ratios of the original field $F$ (dashed lines).
We find that the angularly averaged function has a global maximum at 
$q_1^{(1)} = - 0.261$, where $\bar{R}_{g}= f(q_{1}^{(1)}) = 0.1147$, and the global
minimum is at $q_1^{(2)} = - 1.22$, where the signal drops close to zero. 
The zero signal is due to the fact that the signal scales as 
($q_{1}b_{F}^{\delta} + b_{F^{2}}^{\delta}$), and with 
$b_{F}^{\delta}$, $b_{F^{2}}^{2}$ fixed there is always a $q_1$ at which 
the signal vanishes. 
Similar argument applies to the higher order case shown in figure \ref{fig:f3}.

In the previous section we saw that when the flux field is 
Gaussianized, the S/N gain with respect to that of the original flux field
varies anisotropically. 
Interestingly, this is not the case for a generic quadratic transformation, 
since the three gains in figure \ref{fig:f2} are remarkably similar.
One may wonder why. The angular dependence of the signal is set by the redshift space distortion 
parameter of the field, $\beta_g = f b_g^\eta/ b_g^\delta$. 
In our simulations we find that $\beta_{F^2} \sim \beta_F$, and therefore
any linear combination of these fields will have also a similar $\beta_g$, which explains the weak
angular dependence of the S/N gain.

Finally, we note that even at the global maximum $q_1^{(1)}$, the 
predicted signal-to-noise ratio is only $3.40\%$ larger than that of the original field. 
To test this prediction, we have applied the transformation 
$g_2(F) = - 0.261 F + F^2$ to the flux field in the simulations and have measured 
the statistics of the transformed field. 
The measured signal-to-noise ratio is $(3.56 \pm 8.74) \%$ larger than the reference
value, consistent with our prediction. 
The uncertainty in the quoted ratio is set by the uncertainty in our 
measurement of the bias parameters.

\subsubsection{Results at third order}

We consider now generic cubic transformations $g(F)=q_1 F + q_2 F^2 + F^3$. 
After measuring the bias parameters for $F^3$, its 1D power spectrum and 
its cross-power spectra with $F$ and $F^2$, we can extend the previous study
on quadratic fields to include transformations of order $n=3$. 

\begin{figure}
 \centering
 \includegraphics[width = 15cm, height = 15cm,keepaspectratio= true, trim = 1cm 0cm 3cm 0cm]{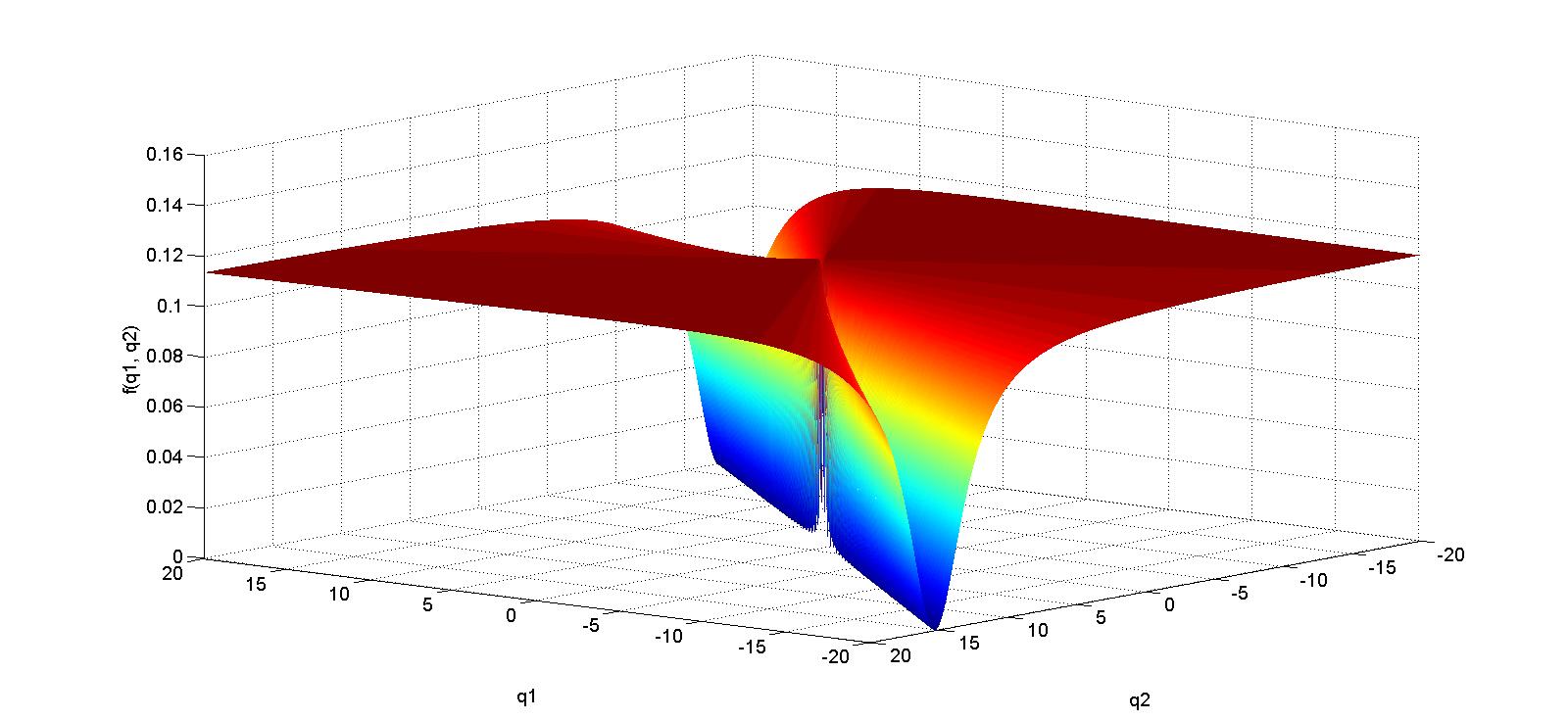}
 \caption{Angularly averaged signal-to-noise ratio for generic cubic 
  transformations, as a function of $q_1$ and $q_2$. }
 \label{fig:f3}
\end{figure}

In figure \ref{fig:f3} we show the angularly averaged signal-to-noise ratio 
for generic cubic transformations. 
The global maximum corresponds to the transformation 
$g_3(F)=1.084F - 3.039F^2 +F^3$, where $\bar{R}_{g}$ is only $3.47\%$ larger than
that of the original $F$ field. We have also tested this in simulations, by
applying the transformation $g_3(F)$ to the flux field and measuring its
statistics. The gain measured in simulations is $(4.09 \pm 8.79) \%$.

\begin{table}
 \begin{tabular}{|c||c|c|c|c|c||c|c|c|}
  \hline
$g$-field & $\langle g \rangle $ & $b_g^\delta$ & $b_g^{\eta}$ & $\beta_g$ & $P_g^{1D}(k_0)$ & $\overline{R}_{g}$ & $R_g(\mu{=}0)$ & $R_g(\mu{=}1)$ \\  \hline

$F$ & 0.8413 & -0.0733 & -0.0994 & 1.302 & 0.1101 & 0.1109 & 0.0487 & 0.2710 \\ \hline
$F^2$ & 0.7703 & -0.0902 & -0.1194 & 1.271 & 0.1587 & 0.1145 & 0.0512 & 0.2769 \\ \hline
$F^3$ & 0.7197 & -0.0994 & -0.1248 & 1.205 & 0.1861 & 0.1143 & 0.0531 & 0.2701 \\ \hline
$g_2(F)$ & 0.5509 & -0.0712 & -0.0935 & 1.261 & 0.0980 & 0.1148 & 0.0517 & 0.2768 \\ \hline
$g_3(F)$ & -0.7094 & 0.0957 & 0.1302 & 1.306 & 0.1809 & 0.1154 & 0.0507 & 0.2824 \\ \hline
Gaussianized & $0.0080$ & -0.3706 & -0.2679 & 0.694 & 2.1132 & 0.1031 & 0.0650 & 0.1929 \\ \hline
 \hline
 \end{tabular}
 \caption{Summary of values used in this section (measured directly from simulation). For each field $g$, we 
  present its mean $\langle g \rangle$, its density bias $b_g^\delta$, velocity bias 
  $b_g^\eta$, redshift space distortion parameter $\beta_g$, 
  1D power $P^{1D}_g(k_0)$ in low-$k$ limit, angularly averaged signal-to-noise ratio $\bar{R}_g$,
  transverse ratio $R_g(\mu{=}0)$ and line-of-sight ratio $R_g(\mu{=}1)$. The simulation we use
  is not able to determine the correct sign of bias by itself, and therefore for consistency we set bias negative
  for $g(F)$ with positive mean $\langle g \rangle$ and vice versa. In particular, 
  to compute $\beta$ we have used $f=0.96$ for precision. }
 \label{tab:z2}
\end{table}

We have explored higher order transformations, but the results become
numerically very unstable, and we are not able to reproduce the predicted 
$\bar{R}_{g}$ in simulations. For instance, the noise term for $n=4$ involves 
a total of ten 1D auto- and cross-power spectra, each of which has an 
associated numerical uncertainty.
We have also repeated the analyses at different redshift outputs ($z=2.4$ and 
$z=2.75$), finding results qualitatively similar. 

Finally, in table \ref{tab:z2} we list several relevant quantities measured 
in the simulations for some of the fields mentioned in this section, 
including the quadratic and the cubic transformations with higher angularly averaged signal-to-noise 
ratio, as well as the Gaussianized field. Note that we cannot exclude that there may be other
transformations that give a better improvement than the ones we explored here: the fact that 
we do not approach the gains achieved by Gaussianized field for $\mu=0$ by the third order expansion 
suggests that it may be possible to find other transformations that offer large gains, but the 
Taylor expansion approach used here does not seem to find them at the order we are working.

\section{Analytic model for large-scale bias}
\label{sec:model}

Paper \cite{Seljak 2012} presents an analytic model to describe the biasing of 
the \lyaf\ transmitted flux fraction $F=e^{-\tau}$, where the density
bias $b_F^\delta$ (response to large-scale overdensity) and the velocity 
bias $b^\eta_F$ (response to line-of-sight velocity gradient) are expressed as 
\begin{equation}
 b_F^\delta = \langle \frac{dF}{d\delta} \rangle 
	+ \nu_2 \langle \delta \frac{dF}{d\delta} \rangle ~, \qquad
 b_F^\eta = \langle \tau \frac{dF}{d\tau} \rangle ~,
 \label{eq:bF}
\end{equation}
with $\nu_2=34/21$.  A third term is present in the biasing model of \cite{Seljak 2012} that 
is  proportional to the level of non-Gaussianities $f_{NL}$ in the primordial
density fluctuations. In this study we assume that $f_{NL}=0$, and ignore 
this extra term.

The derivation in \cite{Seljak 2012} ignores the shift in the overall scale caused by the long wavelength overdensity. 
For second order $\delta^2$ bias, this shift term comes as $d\ln [k^3P(k)]/6d\ln k$ \cite{Sherwin}, which should 
be compared to $\nu_2=34/21$. 
At the \lyaf\ scale $k \sim 1h/Mpc$ $k^3P(k)$ 
has a slope of approximately 0.6, and if we smooth \lyaf\ with even smaller R (higher k) then it is even smaller. The correction is
0.1 relative to 34/21, i.e. it is of order 6% correction to 
$\delta^2$ bias term. We will also continue to ignore this correction here. 

\subsection{Biasing of a generic transformation}

In section \ref{ss:Fn} we have studied the effect of applying a generic 
analytic transformation to the \lyaf\ transmitted flux fraction 
$g(F)=\sum_{m=1}^\infty a_m F^m$. 
Because the derivation of eq.\ref{eq:bF} in paper \cite {Seljak 2012} can be applied to
an arbitrary analytic function of $\delta$, we can thus compute the bias parameters of the tranformed
 field by replacing $F$  with $g(F)$ in eq.\ref{eq:bF}:
\begin{equation}
 b_g^\delta = \langle \frac{dg(F)}{d\delta} \rangle 
        + \nu_2 \langle \delta \frac{dg(F)}{d\delta} \rangle ~, \qquad
 b_g^\eta = \langle \tau \frac{dg(F)}{d\tau} \rangle ~.
 \label{eq:bg}
\end{equation}
Furthermore, using the chain rule we can substitute the derivatives of $g(F)$ by 
the corresponding derivatives of $F$ multiplied by the derivative of the transformation itself (i.e., 
$ \frac{dg(F)}{dF} = \sum_{m=1}^\infty m a_m F^{m-1} $). 
The biasing of $g(F) $ can then be expressed as 
\begin{equation}
b_g(\mu) = b_g^\delta + (f \mu^2)b_g^\eta
	= \langle (\sum_{m=1}^\infty m a_m F^{m-1}) (\frac{dF}{d\delta} 
		+ \nu_2 \delta \frac{dF}{d\delta}) \rangle 
	+ f \mu^2 \langle (\sum_{m=1}^\infty m a_m F^{m-1}) 
		\tau \frac{dF}{d\tau} \rangle ~.
\label{eq:bg2}
\end{equation}

As pointed out in \cite{Seljak 2012}, the velocity bias $b_F^\eta$ is 
completely determined by the probability distribution function (PDF) 
of the field, $p_F(F)$. 
This is also true for a generic transformation, where the velocity bias 
can be computed as:
\begin{equation}
 b^\eta_g = \sum_{m=1}^\infty m a_m \langle F^m ~ \ln(F) \rangle 
	= \sum_{m=1}^\infty m a_m \int_0^1 dF ~ p_F(F) ~ F^m ~ \ln(F) ~.
\label{eq:bgv}
\end{equation}
The PDF can be computed both in the data and in hydrodynamic simulations,
allowing for a test of the model presented in \cite{Seljak 2012}.

On the other hand, the computation of the density bias $b_g^\delta$ requires
the derivative of the transmitted flux fraction with respect to the density 
field, $\frac{dF}{d\delta}$. 
In order to compute this value we have to assume an analytic relation 
$F(\delta)$, which will strongly affect the predicted value. 
In the next subsection, we will present the predictions for 
$b_g^\eta$ and $b_g^\delta$ using a particular model of $F(\delta)$.

\subsection{Bias computation with Fluctuating Gunn-Peterson Approximation}
\label{ss:FGPA}

The Fluctuating Gunn-Peterson Approximation (FGPA) suggests that the optical
depth of \lya\ has a simple dependence on the local density of the IGM:
\begin{equation}
 \tau = A(1+\delta)^\alpha ~.
 \label{eq:FPGA}
\end{equation}
Here, $\alpha = 2 - 0.7 (\gamma-1) = 2 - 0.7 \frac{d(ln\rho)}{d(lnT)}$, 
where $\rho$ is the intergalactic gas density and $T$ is the temperature of 
the IGM, the amplitude $A$ is proportional to $T^{-0.7} \Gamma^{-1}$ 
where $\Gamma$ is the ionization rate by cosmic UV backgroud 
\cite{Weinberg 2003, Dave 2003}. 
In particular, $\alpha$ typically ranges from 1.6 to 2 
\cite{Weinberg 2003}. 
The FGPA has been shown to be a good approximation at the relevant redshifts of 
interest for BAO measurements $2 \lesssim z \lesssim 3.5$. However,
 a \textit {caveat}: eq.\ref{eq:bF} was derived in paper 
\cite {Seljak 2012} on the basis of Taylor expansion
\begin{equation}
\tau (\delta) = \sum_{n=0}^{\infty} \frac{\tau^{(n)}(0)} {n!} \delta^{n},
\label{eq:4.6}
\end{equation}
where the convergence radius is 1 in FGPA, which implies the application of 
FGPA in our prediction is only valid for at most $|\delta| \leq 1$. For crude test, we do not restrict
ourselves to this domain of convergence in the following study, but we are aware
that this domain restriction of FGPA might be responsible for any discrepancy between
 the predictions and the measurements in section \ref{ss:test}.

Now, differiating $F$ with respect to $\delta$ according to eq.\ref{eq:FPGA} and 
defining the auxiliary function 
$r(F) = \alpha \ln(F) (
	\nu_2 + (1-\nu_2)(-\frac{\ln(F)}{A})^{-\frac{1}{\alpha}})$, 
we can express the density bias of a generic transformation as 
\begin {equation}
 b_g^\delta = \sum_{m=1}^\infty m a_m \langle F^{n} ~ r(F) \rangle
	= \sum_{m=1}^\infty m a_m \int_0^1 dF ~ p_F(F) ~ F^m ~ r(F) ~.
 \label{eq:bgd}
\end{equation}

Finally, to compute eq.\ref{eq:bgv} and eq.\ref{eq:bgd}, besides the PDF from data or from a simulation we can also
use the theoretically approximated PDF in which the density field $\delta$ follows a log-normal 
distribution:
\begin{equation}
1 + \delta = e^{(\delta_G - \frac{\sigma^2}{2})} ~, 
	\qquad p_G(\delta_G) = \frac{1}{\sqrt{2\pi \sigma^2}} 
		e^{-\frac{\delta_G^2}{2\sigma^2}} ~,
\label{eq:lognorm}
\end{equation}
where $p_G(\delta_G)$ is the PDF of the auxiliary Gaussianized field $\delta_G$.

\subsection{Testing the model with simulations}
\label{ss:test}

In order to use the FPGA and the log-normal model for the density field, 
we need to choose a value for the parameters $\alpha, A,$ and $\sigma$. 
Using the hydrodynamic simulations presented in section \ref{ss:sims}, 
we fit these parameters by comparing the model predictions and the simulation measurements
of three different statistics: flux PDF $p_F(F)$, flux 
moments $\langle F^n \rangle$ and $\langle F^{n}lnF \rangle $. 

\begin{figure}
 \centering
 \includegraphics[width = 17cm, height = 13cm,keepaspectratio= true, trim = 2cm 17.8cm 0cm 3.3cm]{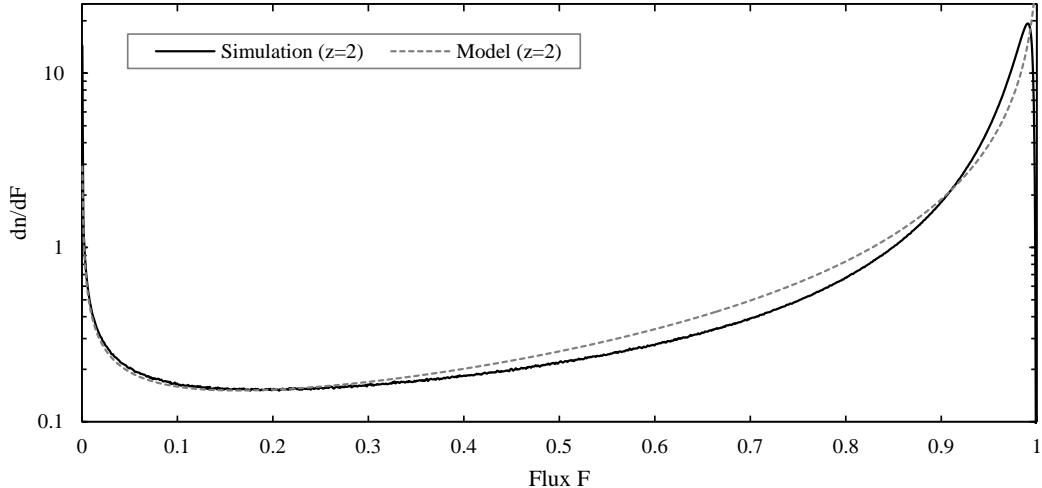}
 \caption{Comparison of the flux PDF measured in simulations with that 
  predicted by the best fit parameters of the FGPA + log-normal model.}
\label{fig:pdf}
\end{figure}

In this section we will focus on the results at $z=2$, but we study other 
redshift outputs in appendix \ref{app:z}. The results are qualitatively 
similar, although in general we find the fits at lower redshift to be slightly
better.
In figure \ref{fig:pdf} we compare the flux PDF measured in the simulation
with that predicted by the best fit model: $\alpha = 1.22$, $A = 0.31$ 
and $\sigma = 1.72$. 
Notice that the best fit value of $\alpha$ is smaller than the typical range 
of $1.6 < \alpha < 2$.

\begin{figure}
 \centering
 \includegraphics[width = 17cm, height = 13cm,keepaspectratio= true, trim = 2cm 17.8cm 0cm 2.9cm] {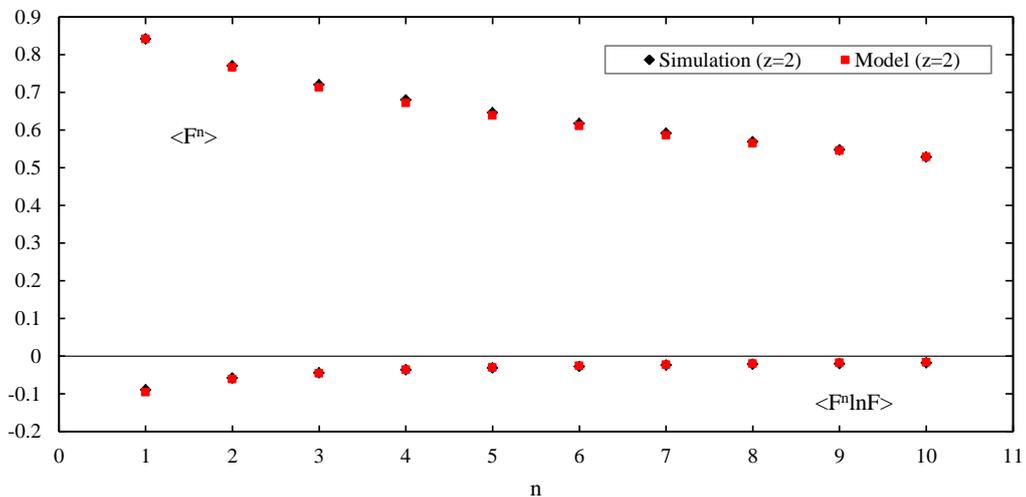}
 \caption{Comparison of the first flux moments in simulations with those
  predicted by the best fit parameters of the FGPA (upper panel),
  and of the averages of $\langle F^n \ln(F) \rangle$ relevant to the computation of velocity
  biases (lower panel).}
\label{fig:Fn}
\end{figure}

Even though the first moments $\langle F^n \rangle $ are quite well
fitted, as shown in figure \ref{fig:Fn}, it is clear from figure \ref{fig:pdf}
that no combination of parameters is able to reproduce the measured PDF, 
especially at the high flux end. 
Qualitatively, we find the following trends:
(1) increasing $\alpha$ moves the moments $\langle F^{n} \rangle$ 
\textit {up} and vice versa, but higher moments are more sensitive to the 
change of $\alpha$ than lower moments with the first moment 
$\langle F \rangle$ being almost immune to the change of $\alpha$;  
(2) increasing parameter $A$ readily moves all the moments 
$\langle F^{n} \rangle$ \textit {down} by similar amount, and vice versa; 
(3) changing $\sigma$ has similar effect on the moments as changing $\alpha$, 
but the moments (especially the lower moments) are generally more sensitive 
to the change of $\sigma$ than in $\alpha$ case.  

As described above, in order to predict the value of the bias parameters we 
need to use a flux PDF. Since the best fit model does not reproduce very well
the PDF measured in simulations, we will present two predictions for the bias
parameter: one using the \textit{model PDF} and one using 
the \textit{simulation PDF}.

\begin{figure}
 \begin{subfigure} {.5\linewidth}
 \includegraphics[width = 9cm, height = 15cm,keepaspectratio= true, trim = 3.5cm 15cm 0.6cm 2.8cm]{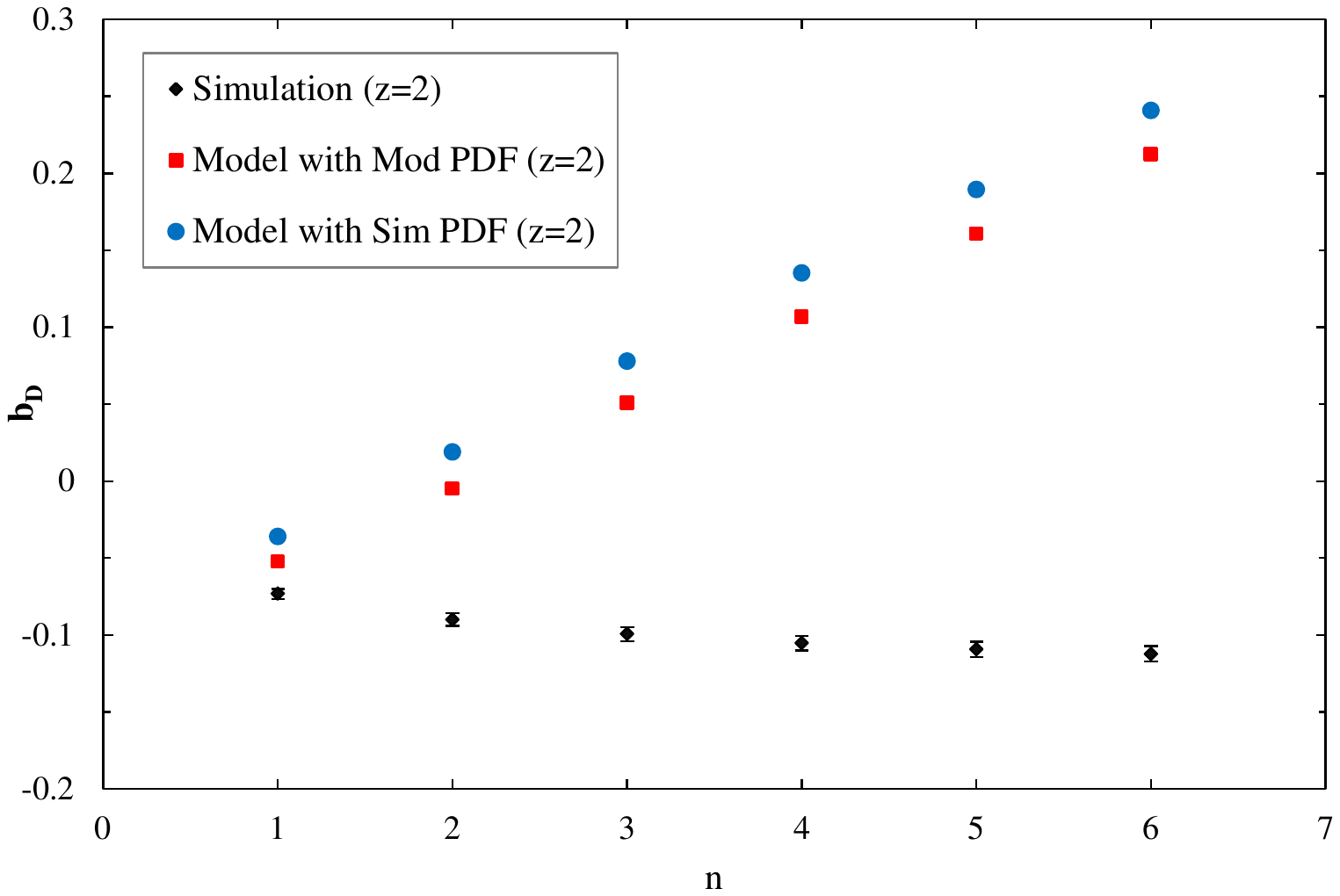}
 \caption{Density bias $b^\delta_{F^n}$.}
 \label{fig:sfig1}
\end{subfigure}
\quad
\begin{subfigure} {.5\linewidth}
 \includegraphics[width = 9cm, height = 15cm,keepaspectratio= true, trim = 3.6cm 15cm 0.5cm 2.8cm]{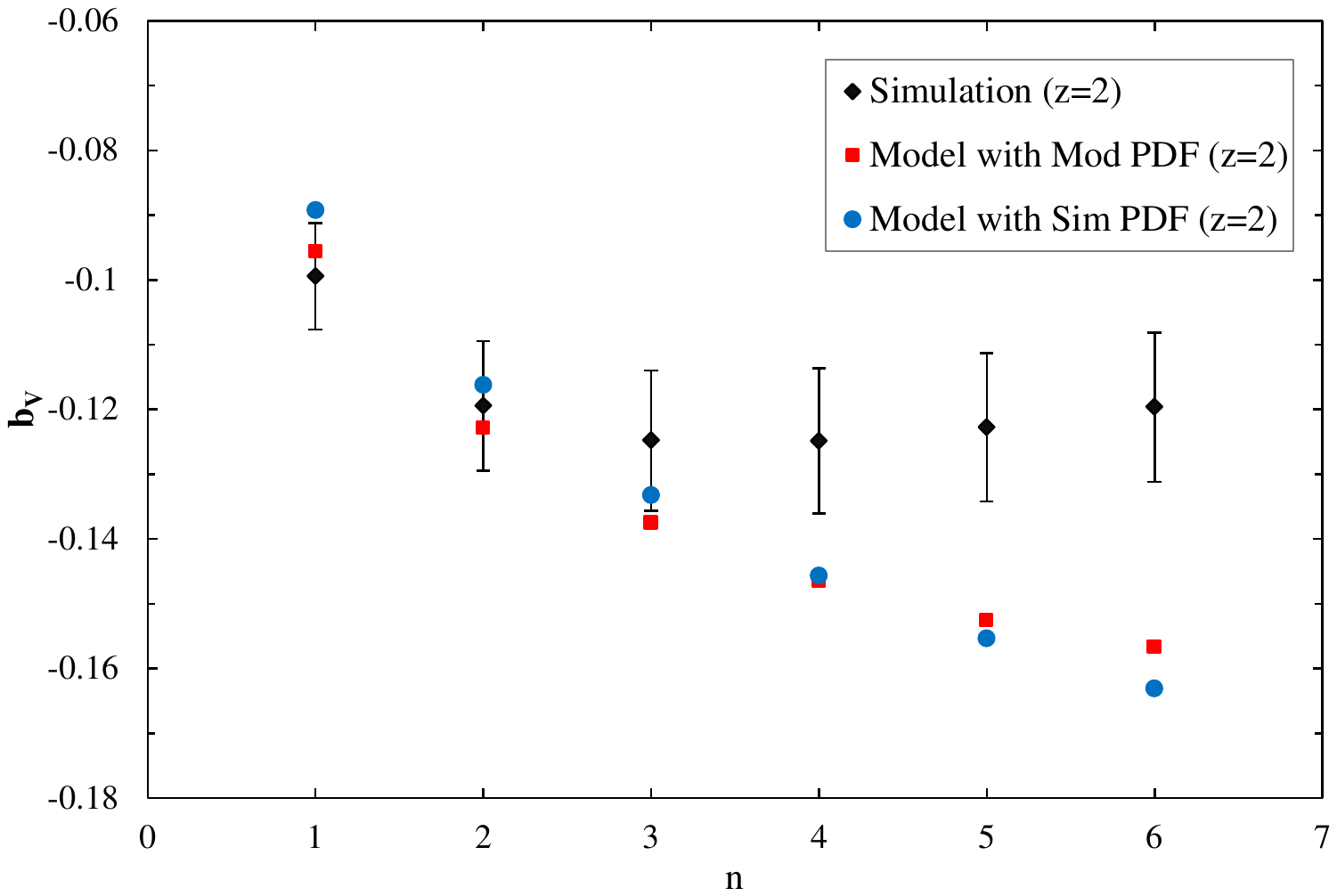}
 \caption{Velocity bias $b^\eta_{F^n}$.}
 \label{fig:sfig2}
\end{subfigure}
\caption{Comparison of predicted and measured biases at $z=2$. 
 The blue and the red data points in both panels correspond to the parameter 
 best-fitting values presented in the main text.}
\label{fig:bFn}
\end{figure}

In figure \ref{fig:bFn} we show the predictions for the bias parameters of
$F, F^2, \ldots, F^6$, compared to the values measured in simulations at 
$z=2$, where the prediction using the model PDF is somewhat closer than the 
prediction using the measured PDF \textendash\ however, as shown in appendix \ref{app:z},
 this is not the case in other redshifts.  From the plots it is clear that while we are able 
to correctly predict the 
velocity bias for the first orders ($b^\eta_F, b^\eta_{F^2}, b^\eta_{F^3}$), 
the model fails to reproduce the measured density bias even for the original
field $b_F^\delta$. In fact, it is not surprising that the prediction for the density bias is
 worse than the prediction for the velocity bias, since the former not only depends on 
the PDF but also in the assumed relation $\tau(\delta)$. And as discussed in section
\ref{ss:FGPA} that the FGPA application is restricted by $|\delta| \leq 1$, this domain
restriction may be responsible for the poor predictions of density bias.

\begin{figure}
 \centering
 \includegraphics[width = 17cm, height = 13cm,keepaspectratio= true, trim = 2cm 17.8cm 0cm 2.9cm]{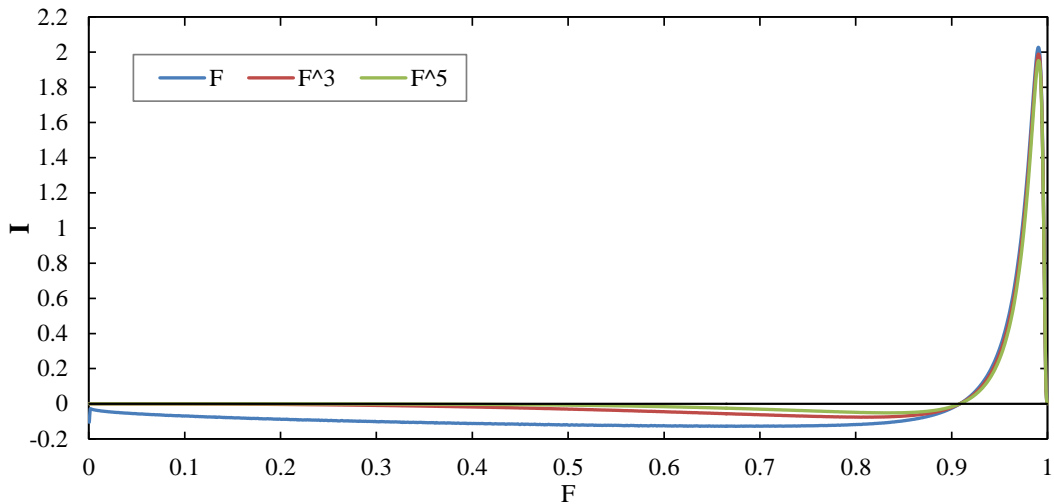}
 \caption{Plots of integrands $I = p_{F}(F) F^{n}r(F)$ in eq.\eqref{eq:bgd} for 
  density bias $b_{F}^{\delta}, b_{F^{3}}^{\delta} $ and $b_{F^{5}}^{\delta}$, where $p_{F}(F)$ is the 
  simulation flux PDF.}
 \label{fig:int}
\end{figure}

The prediction of the density bias for the higher order fields $F^n$ is 
particularly difficult, since it is completely dominated by the high $F$ 
end of the PDF, where the model performs poorly. This is clearly seen 
in figure \ref{fig:int}, where we show the integrand of equation \ref{eq:bgd} 
for three fields of interest: $F$, $F^3$ and $F^5$.

\section {Conclusions}
\label{sec:conc}

In the context of optimizing the measurement of the BAO scale in \lyaf\ surveys,
we have presented a study of the effect of non-linear transformations of the
transmitted flux fraction $F \rightarrow g(F)$ on the expected signal-to-noise ratio of 
the measurement. 

On the large scales relevant in a BAO measurement, the signal is proportional 
to the square of the large-scale biasing of the field. 
The noise, on the other hand, has several contributions. 
In the limit of being dominated by \textit{aliasing noise} (equivalent to 
shot-noise in a galaxy survey), the noise is proportional to the amplitude of 
the one-dimensional power spectrum at low-$k$ limit. 
Under these assumptions, we study the signal-to-noise ratio for different
transformations using hydrodynamic simulations.

The first transformation that we have studied is a monotonic relation $g(F)$
so that the final field has a Gaussian PDF. 
We have shown that the Gaussianized field obtained with this transformation would 
have a $\approx 33\%$ larger signal-to-noise ratio in the transverse BAO 
measurement, but a $\approx 30\%$ smaller ratio along the line of sight.
The anisotropy of the gain can be explained by a significantly lower value 
of the redshift space distortion parameter $\beta$ in the Gaussianized field. 

We have also studied the case of analytic transformations of the form
$g(F) = \sum_{m=1}^{n} a_m F^m$, and presented results for generic quadratic
($n{=}2$) and cubic ($n{=}3$) transformations. 
In both cases, the angularly averaged maximum gain in signal-to-noise that we measure is 
rather small ($< 5\%$). 

Finally, we have extended the biasing model presented in \cite{Seljak 2012} 
to describe the biasing of higher order field $F^n$. 
We have shown that while the model is able to describe the velocity bias
$b^\eta$ reasonably well for the first orders ($n{=}1,2,3$), the density bias
$b^\delta$ is difficult to predict even for the original field $F$. 

Our findings may be of use in attempts to optimize BAO signal to noise in 
realistic \lyaf\ surveys.
It is possible that one can use transformations such as Gaussianization to significantly 
improve the measurement of angular diameter distance $D_A$. One could perhaps envision 
that the measurement of $H$ along the line of sight is done using flux field itself, while 
measuring $D_A$ in the transverse direction would be done using Gaussianized field. 
For a more realistic estimate of the actual gains 
one should include measurement noise and resolution in the analysis, which was ignored here. 
These will depend on individual surveys and are beyond the scope of this paper, but should 
be explored further in the future. 

\acknowledgments {This work is supported partly by the 2013 Summer 
Undergraduate Research Fellowship Program (SURF) at the University of 
California, Berkeley, and by NASA ATP grant NNX12AG71G. 
We would like to thank Nishikanta Khandai for providing the hydrodynamic 
simulation, and An\v{z}e Slosar, Patrick McDonald and Martin White for 
useful discussions.}

\appendix

\section{Bias fitting procedures}
\label{app:b}

The first step towards measuring the bias parameters of the \lya\ 
transmission flux fraction $F$ is to measure its 3D power spectrum. 
We measure band powers in a grid of 3 wavenumber bins limited by 0.0, 0.4, 0.8 
and 1.2 $\ihMpc$, and in 4 bins in angular coordinate 
$\mu = k_\parallel / k$, limited by 0, 0.25, 0.5, 0.75 and 1, 
for a total of $N=12$ bands. 
 
We start by adding the norm of the relevant 3D Fourier modes 
(obtained with a FFT) within each of the $N$ bands, and treat these values as 
our data vector (or $\vPo$). 
We then add the prediction from our fiducial model (described below) for each 
of the modes in the bands, and treat these as the theoretical prediction 
(or $\vPt$). 
We estimate the (diagonal) covariance $\vC$ of the data vector by adding 
$2 \vPt^2$ for each mode within a band. 
We use these ingredients to compute a maximum likelihood estimator of the band power, 
where the likelihood $L$ is proportional to:  
\begin{equation}
L \propto \det\left(\vC\right)^{-1/2}
\exp\left[-\frac{1}{2} \left(\vPo-\vPt \right)^t \vC^{-1} 
	\left(\vPo -\vPt \right) \right] ~.
\end{equation}

We are interested in measuring the scale-independent density bias $b^\delta$ 
and the velocity bias $b^\eta$ parameters, which describe the clustering on 
linear scales. 
However, it is difficult to fit linear bias parameters in a box that is 
only 40 $h^{-1}$Mpc wide, since there are very few Fourier modes that are 
in the linear regime. 
Therefore, we need to model the deviations from linear theory in the 
simulations, and marginalize over the free parameters these. 
Based in the description of large scale biasing of \cite {McDonald 2009}, 
we parameterize the deviations from linear theory with the following analytic 
model: 
\begin{equation}
P_F (k, k_\parallel) = \Bigg[ b^\delta(k, k_\parallel) 
	+ b^\eta(k, k_\parallel) f \frac{k_\parallel^2}{k^2} \Bigg]^2 P(k) 
	+ N(k, k_\parallel),
\end{equation}
with
\begin{equation}
b^\delta (k, k_\parallel) = b_\delta^{0, 0} + b_\delta^{2, 0} (R k)^{2} + b_\delta^{0, 2} (R k_\parallel)^{2} + b_\delta^{4, 0}(R k)^{4} + b_\delta^{0, 4} (R k_\parallel)^{4} + b_\delta^{2,2} (R k)^{2} (R k_\parallel)^{2},
\end{equation}
where $b^{m,n}$ are free parameters, and the equivalent for 
$b^\eta(k,k_\parallel)$ and $N(k, k_\parallel)$. 
In the expressions above, $R=0.2h^{-1}$ Mpc is a typical non-linear scale, 
and $P(k)$ is the linear matter power spectrum in the simulation. 
In total, we have 18 free parameters, including 6 parameters describing the 
(shot) noise in the measurement, but we are only interested in the values 
of $b^\delta$ and $b^\eta$ after marginalizing over the other 16 parameters. 
We have made sure that our results do not vary significantly if we remove 
all terms with four order in $R$, or if we increase/reduce the non-linear 
scale by a factor of 2. 

In order to estimate the uncertainties in our measured 3D power, we need 
to have an initial guess of the clustering. 
In practice, we iterate between the measurement of the 3D power with the 
fitting of the bias parameters, until it converges 
(usually within $<$ 5 iterations).

\section {Large-scale bias for z=2.4 and 2.75}
\label{app:z}

In section \ref{sec:model} we presented the predictions for the large scale
bias at redshift $z=2$. 
Here we present some results at redshifts $z=2.4$ and $z=2.75$, which are 
qualitatively similar to those at $z=2$. 

\begin{figure}
\begin{subfigure}{.5\linewidth}
\includegraphics[width = 9cm, height = 15cm,keepaspectratio= true, trim = 3.6cm 15cm 0.5cm 2.8cm]{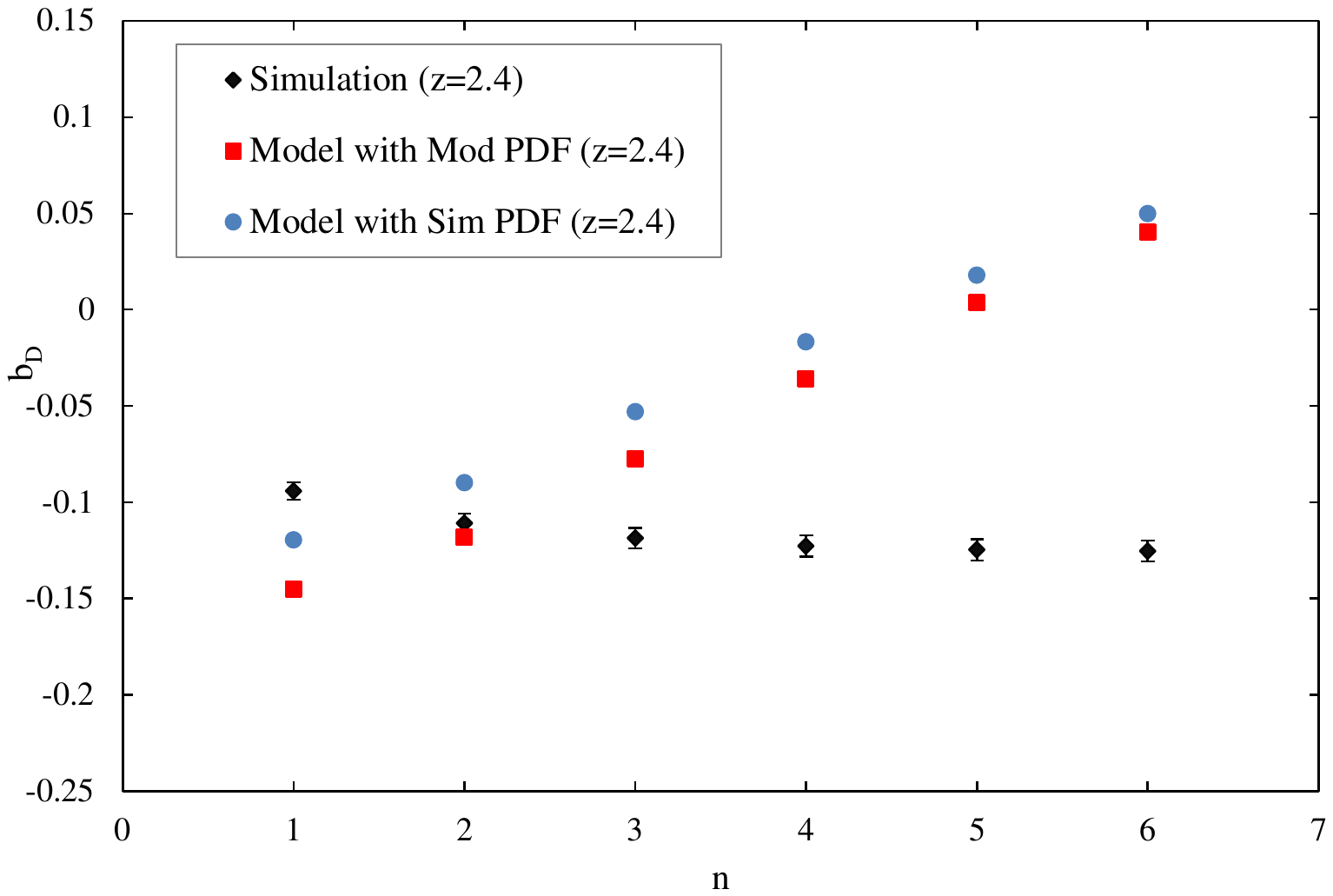}
\caption{Basis density biases $b^\delta_{F^n}$.}
\label{fig:sfig5}
\end{subfigure}
\quad
\begin{subfigure}{.5\linewidth}
\includegraphics[width = 9cm, height = 15cm,keepaspectratio= true, trim = 3.6cm 15cm 0.5cm 2.8cm]{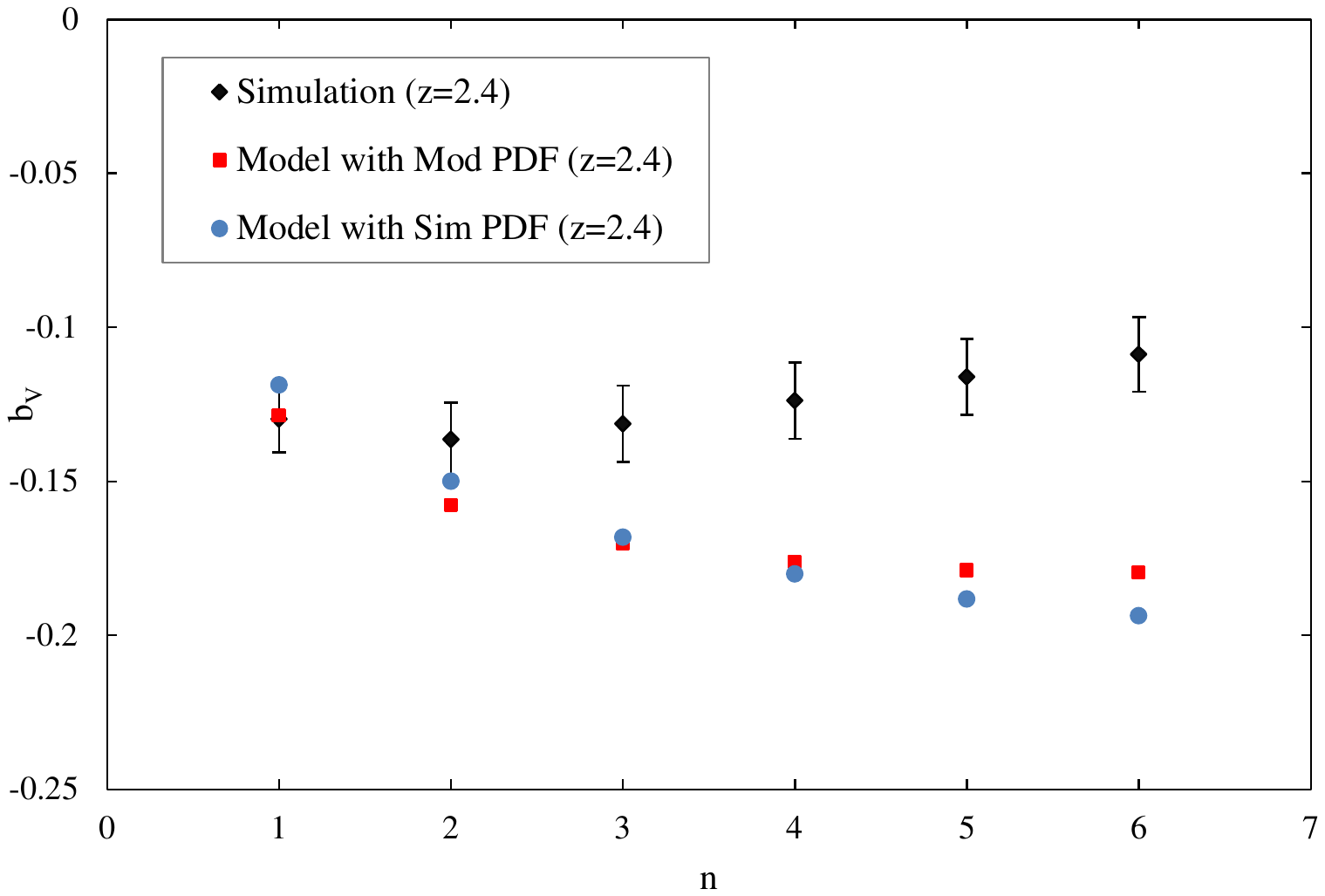}
\caption{Basis velocity biases $b^\eta_{F^n}$.}
\label{fig:sfig6}
\end{subfigure}
\caption{Comparison of predicted and measured basis biases at z=2.4.}
\label{fig:9}
\end{figure}

\begin{figure}
\begin{subfigure}{.5\linewidth}
\includegraphics[width = 9cm, height = 15cm,keepaspectratio= true, trim = 3.6cm 15cm 0.5cm 2.8cm]{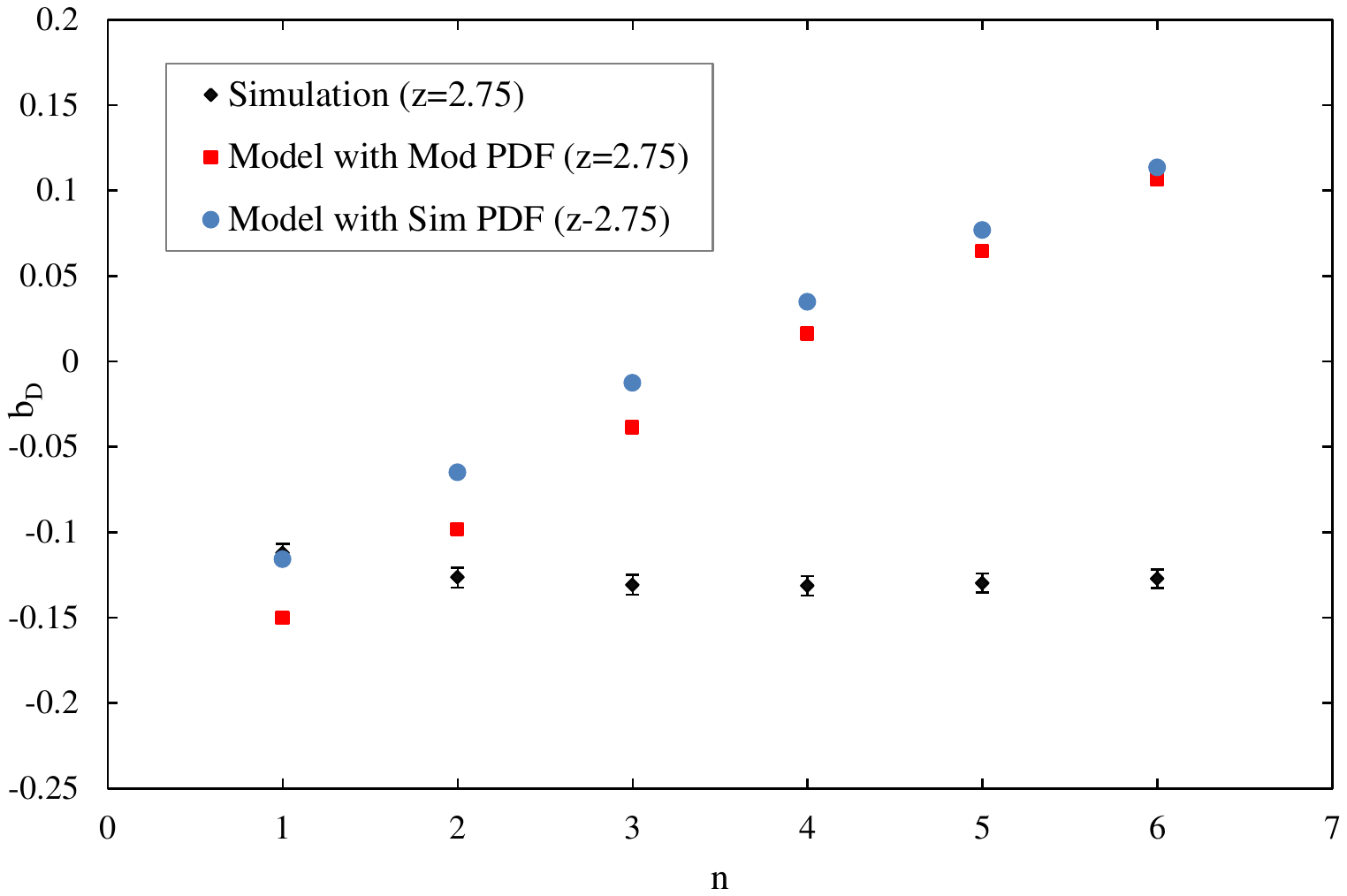}
\caption{Basis density biases $b^\delta_{F^n}$.}
\label{fig:sfig7}
\end{subfigure}
\quad
\begin{subfigure}{.5\linewidth}
\includegraphics[width = 9cm, height = 15cm,keepaspectratio= true, trim = 3.6cm 15cm 0.5cm 2.8cm]{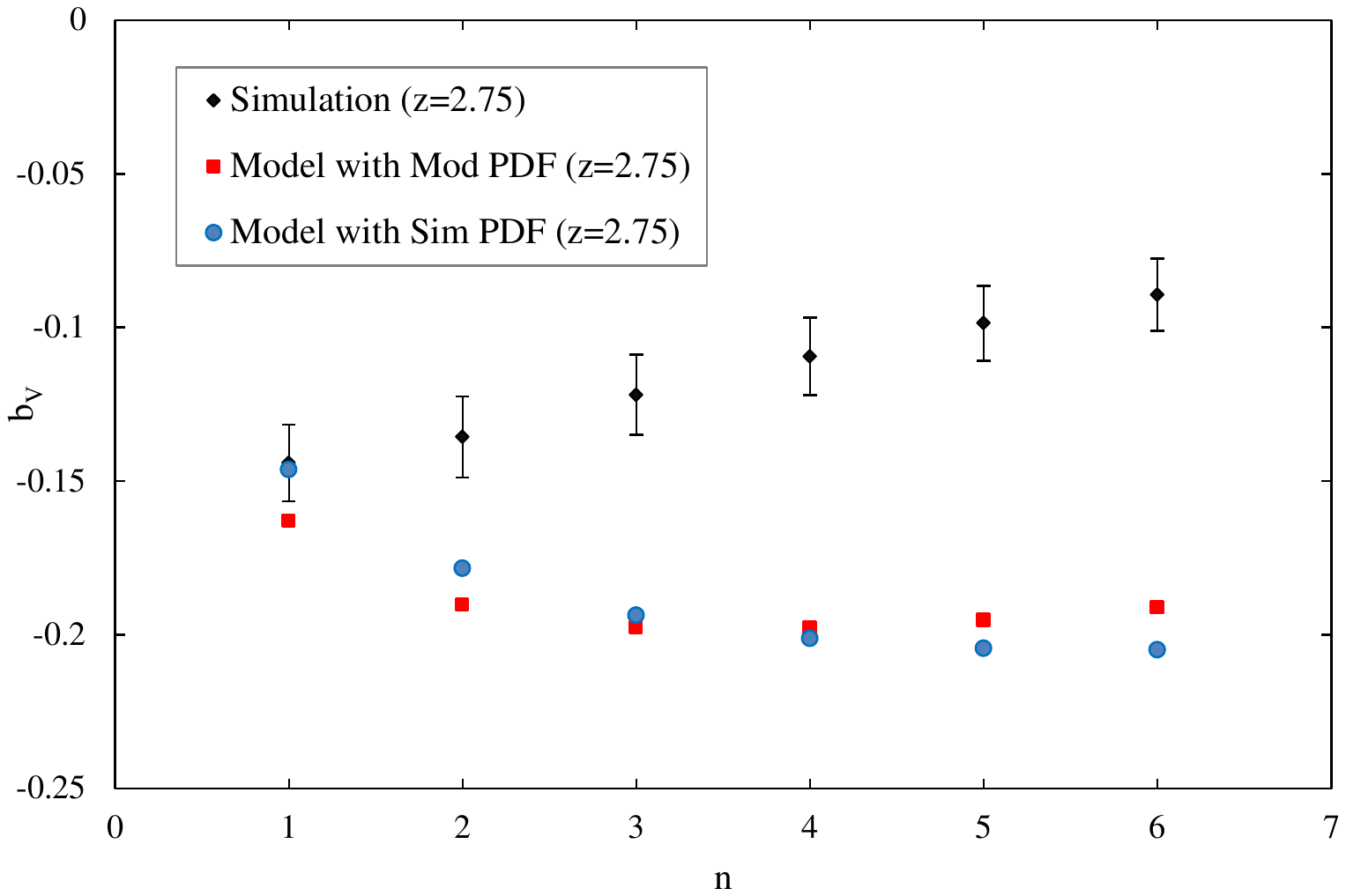}
\caption{Basis velocity biases $b^\eta_{F^n}$.}
\label{fig:sfig8}
\end{subfigure}
\caption{Comparison of predicted and measured basis biases at z=2.75.}
\label{fig:10}
\end{figure}

The parameters in the FGPA are redshift dependent. At $z=2.4$, the best fit
values are $\alpha = 1.465$, $A=0.357$ and $\sigma=1.275$, while at $z=2.75$ 
we find $\alpha = 1.383$, $A=0.511$ and $\sigma = 1.175$. 
At different redshifts, simulation PDF (of flux) performs differently with 
respect to model PDF (of flux) for predicting the biases. 
As we have seen in figures \ref{fig:pdf} and figure \ref{fig:Fn}, at $z=2$ 
model PDF performs better than simulation PDF for both the density biases 
and the velocity biases. 
However, from figures \ref{fig:9} it is hard to distinguish which version of 
PDF works better in general at $z=2.4$, and from figures \ref{fig:10} it is 
obvious that simulation PDF outperforms model PDF at $z=2.75$. 
Overall, observation suggests at high redshift simulation PDF performs better 
than model PDF, and vice versa. 

\end{document}